\documentclass[12pt,twoside]{article}
\usepackage{amssymb}
\usepackage{amsmath}
\usepackage{latexsym}
\usepackage{longtable}
\usepackage{epsfig}
\usepackage{graphicx,bbm,psfrag}
\graphicspath{{images/}}

\begin{document}
\begin{center}
{\Large {\bf Collisional Invariants for the Phonon Boltzmann
Equation\bigskip\bigskip\\}}
{\large{Herbert Spohn}}\footnote{{\tt spohn@ma.tum.de}}\medskip\\
Zentrum Mathematik and Physik Department, TU M\"{u}nchen,\\
D - 85747 Garching, Boltzmannstr. 3, Germany
\end{center}\bigskip\bigskip\bigskip\bigskip
\setcounter{figure}{0}
{\bf Abstract:} For the phonon Boltzmann equation with only pair
collisions we characterize the set of all collisional invariants
under some mild conditions on the dispersion
relation.\bigskip\\

In the study of the Boltzmann equation, collisional invariants play
an important role: for the spatially homogeneous equation they are
in one-to-one correspondence with its stationary solutions and for
the linearized Boltzmann equation they yield the eigenvectors
spanning the zero subspace. In the kinetic theory of gases, under
rather general conditions, a collisional invariant is necessarily of
the form $\psi(v)=a\frac{1}{2}v^2+b\cdot v +c$ with arbitrary
coefficients $a,b,c$ \cite{1}. In case the particles are
relativistic, the kinetic energy, $\frac{1}{2}v^2$, will be replaced
by its relativistic cousin $(1+p^2)^{1/2}$, see \cite{2} for the
characterization of the collisional invariants.

In this note we discuss collisional invariants for the phonon
Boltzmann equation with 4-phonon processes only. The essential
difference to the kinetic theory of gases lies in the fact that the
role of the kinetic energy is taken by the dispersion relation
$\omega(k)$ which is a fairly arbitrary, non-negative function on
$\mathbb{T}^d$, the $d$-dimensional torus of wave numbers.

To keep notation simple we discuss a single band model with the
hypercubic lattice $\mathbb{Z}^d$ as crystal lattice. Amongst the
allowed 4-phonon processes we study first only the number-conserving
ones. For them, under conditions to be specified, a collisional
invariant is necessarily of the form $\psi(k)=a\omega(k)+c$,
$a,c\in\mathbb{R}$. It is then a simple substitution to check
whether the set of collisional invariants is further reduced by
$c=0$ when taking the remaining 4-phonon collisions into account.

We will work in the extended zone scheme. The dispersion relation
$\omega:\mathbb{R}^d\to\mathbb{R}$ then satisfies $\omega\geq 0$ and
is $\mathbb{Z}^d$-periodic, i.e. $\omega(k+n)=\omega(k)$ for all
$n\in\mathbb{Z}^d$. Physically
\begin{equation}\label{0.0}
\omega(k)^2=\sum_{n\in\mathbb{Z}^d}\gamma(n)e^{i 2\pi k\cdot n}
\end{equation}
with $\gamma$ of exponential decay. Thus $\omega$ is real analytic
except at points where $\omega(k)=0$. For multiband models further
points of non-analyticity may occur because of bund crossings.
Therefore we are led to the following.\medskip\\
\textbf{Assumption 1}. Let $\omega:\mathbb{R}^d\to\mathbb{R}$ be
continuous and $\mathbb{Z}^d$-periodic. There exists a manifold
$\Lambda_0 \subset\mathbb{R}^d$ of codimension at least 1 such that
$\omega\in C^2$ on $\mathbb{R}^d\smallsetminus\Lambda_0$. $\omega$ has
bounded second derivatives which may diverge as $\Lambda_0$ is approached.\medskip\\
\textbf{Definition}. A measurable, $\mathbb{Z}^d$-periodic function
$\psi:\mathbb{R}^d\to\mathbb{R}$ is called a \textit{collisional
invariant} if
\begin{equation}\label{1.1}
\psi(k_1)+\psi(k_2)=\psi(k_3)+\psi(k_4)
\end{equation}
for almost every $(k_1,k_2,k_3,k_4)\in\mathbb{R}^{4d}$ under the
constraint that
\begin{equation}\label{1.2}
    k_1+k_2=k_3+k_4\,,
\end{equation}
\begin{equation}\label{1.3}
\omega(k_1)+\omega(k_2)=\omega(k_3)+\omega(k_4)\,.
\end{equation}\medskip\\
\textbf{Assumption 2}. Let
$\Lambda_{\textrm{Hess}}=\{k\in\mathbb{R}^d\setminus\Lambda_0$, det
Hess $\omega(k) = 0\}$. $\overline{\Lambda}_{\textrm{Hess}}$ is a
set
of codimension at least 1.\medskip\\
\textbf{Proposition}. Let $d \geq 2$ and let $\omega$ satisfy Assumptions 1
and 2. Furthermore
\begin{equation}\label{1.4}
\int_{M^\ast} |\psi(k)|dk<\infty
\end{equation}
with $M^\ast=\{k=(k^1,\ldots,k^d)\in \mathbb{R}^d \mid |k^j|\leq
1/2\,,\;j=1,\ldots,d\}$. Then a collisional invariant is necessarily
of the form
\begin{equation}\label{1.5}
\psi=a\omega+c \qquad \textrm{a.s.}
\end{equation}
for some constants $a,c\in \mathbb{R}$.\medskip\\
\textbf{Remark}. The $L^1$-norm in (\ref{1.4}) can be replaced by
any $L^p$ norm, $1\leq p\leq\infty$.\medskip

For the proof we partition $\mathbb{R}^{2d}$ into the sets
$\Lambda_{\eta,\varepsilon}=\{(k_1,k_2)\in\mathbb{R}^{2d}|
k_1+k_2=\eta\,,\;\omega(k_1)+\omega(k_2)=\varepsilon\}$ with
$\Lambda_{\eta,\varepsilon}=\emptyset$ allowed. Let
$\widetilde{\phi} (k_1,k_2)=\psi(k_1)+\psi(k_2)$. Then by assumption
\begin{equation}\label{1.5a}
\int_{(M^\ast)^2} | \widetilde{\phi}(k_1,k_2) | dk_1 dk_2<\infty
\end{equation}
and by definition, except for a set of measure zero,
$\widetilde{\phi}$ is constant on each set
$\Lambda_{\eta,\varepsilon}$.

Let $\phi:\mathbb{R}^d\times\mathbb{R}\to\mathbb{R}$ be smooth and
$\mathbb{Z}^d$-periodic in its first argument. We use the shorthands
$\omega_j=\omega(k_j)$, $\psi_j=\psi(k_j)$,
$\partial^j_\alpha=\partial/\partial k^\alpha_j$,
$\partial_\omega=\partial/\partial\omega$. Then for any test
function $f\in\mathcal{S}(\mathbb{R}^{2d})$ with support away from
$(\Lambda_0\times\mathbb{R}^d)\cup(\mathbb{R}^d\times\Lambda_0)$ one
has
\begin{eqnarray}\label{1.6}
&&\hspace{-10pt}
\int\phi(k_1+k_2,\omega_1+\omega_2)(\partial^1_\alpha-\partial^2_\alpha)
\big(f(\partial^1_\beta-\partial^2_\beta)(\omega_1+\omega_2)\big)dk_1dk_2\nonumber\\
&&\hspace{10pt}=
-\int((\partial^1_\alpha-\partial^2_\alpha)\phi(k_1+k_2,\omega_1+\omega_2))
f(\partial^1_\beta-\partial^2_\beta)(\omega_1+\omega_2) dk_1dk_2\nonumber\\
&&\hspace{10pt}= -\int\partial_\omega\phi(k_1+k_2,\omega_1+\omega_2)
((\partial^1_\alpha-\partial^2_\alpha)(\omega_1+\omega_2))
f(\partial^1_\beta-\partial^2_\beta) (\omega_1+\omega_2) dk_1dk_2\nonumber\\
&&\hspace{10pt}=
-\int((\partial^1_\beta-\partial^2_\beta)\phi(k_1+k_2,\omega_1+\omega_2))
f(\partial^1_\alpha-\partial^2_\alpha)(\omega_1+\omega_2) dk_1dk_2\nonumber\\
&&\hspace{10pt}=\int\phi(k_1+k_2,\omega_1+\omega_2)(\partial^1_\beta-\partial^2_\beta)
\big(f(\partial^1_\alpha-\partial^2_\alpha)(\omega_1+\omega_2)\big)
dk_1dk_2\,,
\end{eqnarray}
integration over $\mathbb{R}^{2d}$. Taking limits the identity
(\ref{1.6}) holds for all $\phi$'s such that
$\int_{(M^\ast)^2}|\phi(k_1+k_2,\omega_1+\omega_2)|dk_1dk_2<\infty$.
Since, by the argument above, $\widetilde{\phi}$ is in this class,
we conclude
\begin{eqnarray}\label{1.7}
&&\hspace{-10pt}\int(\psi_1+\psi_2)\big((\partial^1_\alpha-\partial^2_\alpha)(\omega_1+\omega_2)\big)
(\partial^1_\beta-\partial^2_\beta)f dk_1dk_2\nonumber\\
&&\hspace{10pt}=\int(\psi_1+\psi_2)\big((\partial^1_\beta-\partial^2_\beta)(\omega_1+\omega_2)\big)
(\partial^1_\alpha-\partial^2_\alpha)f dk_1dk_2\,.
\end{eqnarray}

We choose now the particular test function
\begin{equation}\label{1.8}
f(k_1,k_2)=\partial^1_\gamma f_1(k_1)\partial^2_\delta f_2(k_2)
\end{equation}
with $f_1$ and $f_2$ supported away from $\Lambda_0$ and set
\begin{eqnarray}\label{1.9}
&&\hspace{-20pt}
A_{\alpha\beta}(f)=\int\psi(k_1)\partial_\alpha\partial_\beta
f(k_1)dk_1\,,\nonumber\\
&&\hspace{-20pt}
B_{\alpha\beta}(f)=\int\omega(k_1)\partial_\alpha\partial_\beta
f(k_1)dk_1\,,\nonumber\\
&&\hspace{-20pt}A(f_1)=A\,,\;A(f_2)=\widetilde{A}\,,\;
B(f_1)=B\,,\;B(f_2)=\widetilde{B}\,.
\end{eqnarray}
Then, see \cite{3}, Section 12, for more details,
\begin{equation}\label{1.10}
A_{\alpha\gamma}\widetilde{B}_{\beta\delta}+\widetilde{A}_{\alpha\delta}B_{\beta\gamma}
=A_{\beta\gamma}\widetilde{B}_{\alpha\delta}+
\widetilde{A}_{\beta\delta}B_{\alpha\gamma}\,.
\end{equation}

Let us choose $f_1,f_2$ such that $B$ and $\widetilde{B}$ are
invertible and set
\begin{equation}\label{1.11}
C=A B^{-1}\,,\;\widetilde{C}=\widetilde{A}\widetilde{B}^{-1}\,.
\end{equation}
Then, see \cite{3} Appendix 18.4,
\begin{equation}\label{1.12}
C_{\alpha\gamma}\delta_{\beta\delta}+\widetilde{C}_{\alpha\delta}
\delta_{\beta\gamma}= C_{\beta\gamma}
\delta_{\alpha\delta}+\widetilde{C}_{\beta\delta}\delta_{\alpha\gamma}\,.
\end{equation}
Setting $\alpha=1,\beta=2$ and $\gamma=1,2$, $\delta=1,2$, yields
\begin{equation}\label{1.13}
C_{21}+\widetilde{C}_{21}=0\,,\quad C_{12}+\widetilde{C}_{12}=0\,,
\end{equation}
\begin{equation}\label{1.14}
C_{11}=\widetilde{C}_{22}\,,\quad\widetilde{C}_{11}=C_{22}\,.
\end{equation}
In (\ref{1.13}) we choose $f_1=f_2$ to infer that $C_{12}=0$,
$C_{21}=0$. From (\ref{1.14}) we deduce that there is a constant $a$
such that $C_{11}=a$, $C_{22}=a$, independent of the admissable test
function. Repeating for further pairs of indices one concludes that
\begin{equation}\label{1.15}
C(f)=a \mathbbm{1}
\end{equation}
and hence
\begin{equation}\label{1.16}
A(f)=a B(f)
\end{equation}
for all test functions $f$ supported away from $\Lambda_0$ and such
that $B(f)$ is invertible. Since the matrix
$\{\partial_\alpha\partial_\beta
\omega(k)\}_{\alpha,\beta=1,\ldots,d}$ is invertible for $k$ away
from $\Lambda_\textrm{Hess}\cup\Lambda_0$ and since both sets have a
codimension larger than 1, by taking limits, (\ref{1.16}) holds for
all $f\in\mathcal{S}(\mathbb{R}^d)$, to say the collisional
invariant $\psi$ has to satisfy
\begin{equation}\label{1.17}
\int\psi\partial_\alpha\partial_\beta
fdk=a\int\omega\partial_\alpha\partial_\beta fdk
\end{equation}
for all $f\in\mathcal{S}(\mathbb{R}^d)$. Integrating yields
\begin{equation}\label{1.18}
\psi(k)=a\omega(k)+b\cdot k+c\qquad \textrm{a.s.}\,.
\end{equation}
To have $\psi$ periodic forces $b=0$, which is the assertion of the
Proposition.\medskip\\
{\bf Acknowledgement}. I thank Jani Lukkarinen for helpful discussions.

\end{document}